\documentclass[12pt]{article}
\usepackage{amsmath}
\usepackage{amssymb}
\usepackage{graphicx,psfrag,epsf}
\usepackage{enumerate}
\usepackage{natbib}
\usepackage[table,xcdraw]{xcolor} 
\usepackage{pdfpages}
\usepackage{url} 

\usepackage{listings}
\newcommand{\blind}{0}

\begin{document}

\def\spacingset#1{\renewcommand{\baselinestretch}%
{#1}\small\normalsize} \spacingset{1}


\if0\blind
{
  \title{\bf Game time: \\ statistical contests in the classroom}
  \author{Sam Doerken\hspace{1cm} \\
  Institute of Medical Biometry and Statistics, \\
  Faculty of Medicine and Medical Center - University of Freiburg, Germany \bigskip \\
  and\hspace{1cm} \bigskip \\
  Martin Schumacher\hspace{1cm} \\
  Institute of Medical Biometry and Statistics, \\
  Faculty of Medicine and Medical Center - University of Freiburg, Germany \bigskip \\
  and\hspace{1cm} \bigskip \\
  Franz Baumdicker\hspace{1cm} \\
  Department of Mathematical Stochastics, University of Freiburg, Germany}
  \maketitle
} \fi

\if1\blind
{
  \bigskip
  \bigskip
  \bigskip
  \begin{center}
    {\LARGE\bf Game time: \bigskip \\ statistical contests in the classroom}
  \end{center}
  \medskip
} \fi


\bigskip
\begin{abstract}

We describe a contest in variable selection which was part of a statistics course for graduate students. In particular, the possibility to create a contest themselves offered an additional challenge for more advanced students. Since working with data is becoming more important in teaching statistics, we greatly encourage other instructors to try the same.
\end{abstract}

\noindent%
\vfill

\newpage
\spacingset{1.45} 
\section{Introduction}
\label{sec:intro}

Nowadays, teaching statistics through data application is essential \citep{ASA2014, Cobb2015, Horton2015, Wagaman2016}. The ASA guidelines for undergraduate programs in statistical science emphasize the importance for students to have data applications as part of their curriculum \citep{ASA2014}. Similarly, curriculum guidelines for teaching data science also stress that working with data is key \citep{DeVeaux2017}. In his discussion on statistics education, \cite{Cobb2015} puts forth five imperatives for successful teaching, of which he considers the most important to be ``teaching through research''.

Not only is data work beneficial for students, it's also an excellent opportunity to make the learning experience interesting and engaging. The idea to motivate students to learn statistics by creating a fun experience is certainly not new \citep{Goldstein1969, Selkirk1973}, but novel and creative ways of teaching are always well-received by students: \cite{Gerds2016} teaches survival probabilities of the Kaplan-Meier estimate by students performing a ride on the Titanic when it sinks. As students ``drown'', individual survival times are generated by measuring how long students can hold their breath, and from the collective data, the Kaplan-Meier estimate is calculated. \cite{Wierman2016} describes how he uses jokes in the classroom to teach statistics, which students find to be very appealing. As data science courses are quickly gaining popularity, \cite{Hardin2015} point out that data applications can be exciting for students and relevant to their interests.

In a graduate course taught at a university, we sought creative ways to engage students to work with data and make the experience challenging and motivating. The overall course goal was to introduce the theoretical basis of concepts and methods in statistical learning alongside with practical exercises on real and simulated datasets. That meant students should not only gain a theoretical background, but also learn how to distinguish between appropriate and inappropriate methods and assess the performance of a method. The theoretical aspects of the course were mainly covered by the textbook \textit{The Elements of Statistical Learning} \citep{Hastie2009} and the accompanying \textit{An Introduction to Statistical Learning} \citep{James2013} is an excellent source for the corresponding practical exercises in R. Topics of the course included: regression, classification, model assessment and selection, regularization, random forests, neural nets and unsupervised clustering.
 
As  part of the coursework we implemented several exercises in form of a contest, where teams suggest different methods and compete against each other. For an example of such a contest, see also \cite{Pers2009}. This format has become very popular in recent years and there is a growing global community comparing their methods on the platform Kaggle (www.kaggle.com), where participants can compete in predicting and analyzing datasets \citep{Martinez2014}. When analyzing a datasets, it is difficult to know beforehand which technique will be most effective, but the broad spectrum of strategies used in Kaggle competitions has led to many successful models \citep{Taieb2014, Narayanan2011, Lloyd2014}.

The goal of the classroom contests was to motivate students by the competition between teams and rewarding symbolic prizes for the best efforts. In addition, one of the graduate participants suggested to create a classroom competition, related to his research. As part of the coursework he implemented an exercise in variable selection in the form of a contest. This contest strongly contributed to the learning experience, since it allowed students of different levels to gain valuable insights within the same course. In this article we would like to describe the created contest, share our experience and encourage other instructors of statistics courses to integrate games and contests into their course.

\section{The contest}
\label{sec:the_contest}

The objective of the contest was fairly simple. Given a dataset with 20 independent binary variables $X$ and a binary outcome variable $y$, students had to use selection methods in order to identify variables which had an influence on the outcome. The dataset was simulated and the simulation mechanism was kept from the students. Selections made by contestants were rated based on number of true positive and true negative selections using a scoring rule that was previously announced.

The simulated dataset was inspired by the case-control study by \cite{Orriols2010} who investigated the relationship between drug prescriptions (risk factors $X$) and road traffic accidents (outcome $y$) in a large cohort of about 70,000 French drivers between 2005 and 2008. Students were given a presentation on the real dataset to learn about the background and context of the problem. To make the contest computationally more feasible for students, the simulation comprised only 4,000 observations (2,000 cases and controls each) with 20 risk factors of interest. All variables were binary, with 1 indicating that a drug was prescribed to a driver at the time of the accident and 0 indicating that it was not.

A main challenge of the case-control study by \cite{Orriols2010} was the low prevalence of risk factors, meaning that many of the investigated drugs were prescribed only to a small fraction of drivers. Accordingly, prevalences in the simulation ranged from 3\% (1 in 33.3) to 0.1\% (1 in 1,000) uniformly on a log scale, making the data very sparse. Correlation between drugs was kept low.

\begin{figure}
\centering
\fbox{\includegraphics{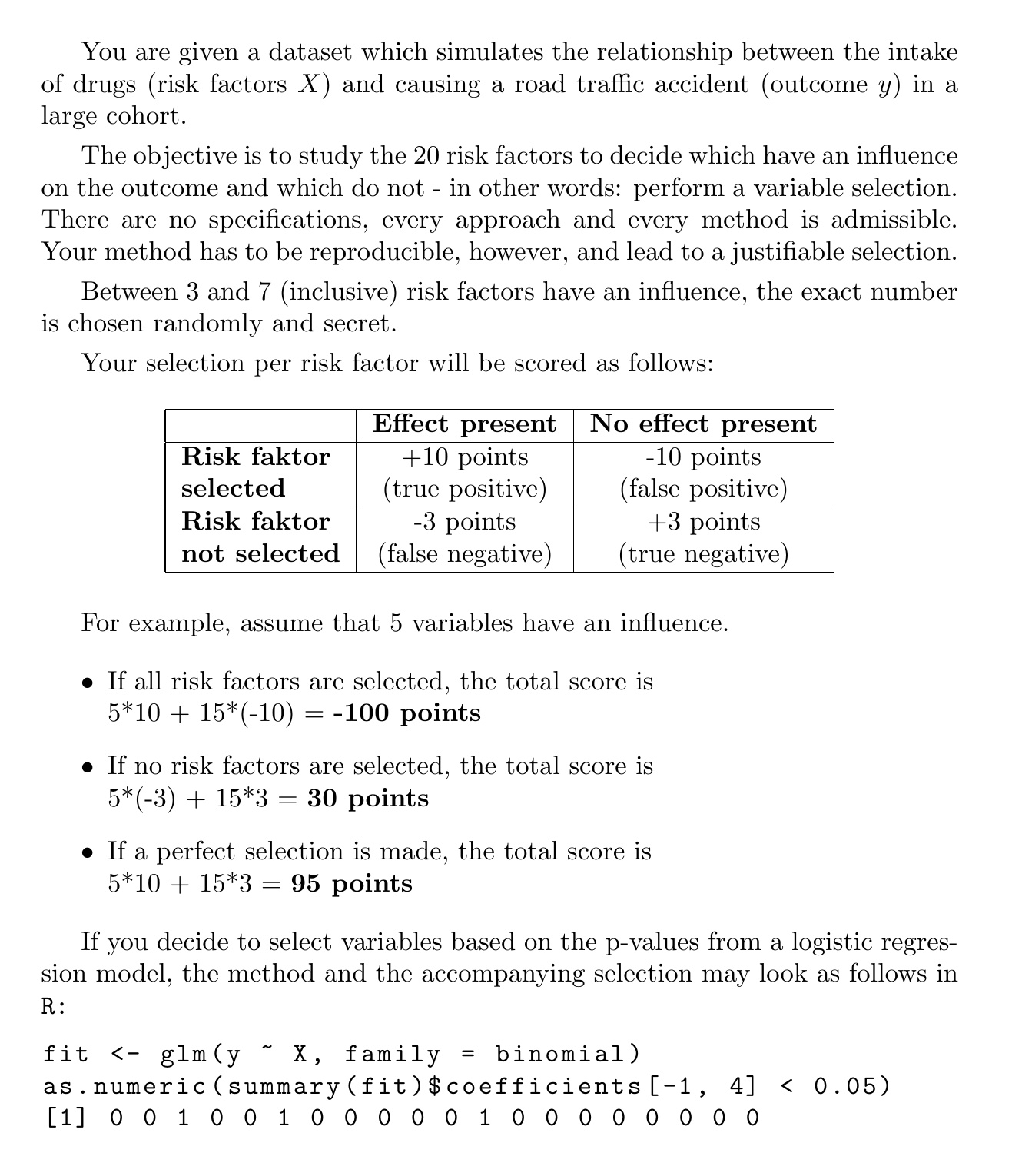}}
\caption{Task description \label{fig:task_description}}
\end{figure}

Ideally when studying a dataset, there is some subject-matter knowledge involved. To imitate this in the contest, it was decided to disclose to the students that the number of risk factors with an influence on the outcome would be randomly between 3 and 7 inclusive. In a real setting, an investigator might have some sense about which results are plausible and which are not. Giving students a range of how many variables could have an influence also allows them to judge the plausibility of their submissions. The instruction given to students is shown in figure \ref{fig:task_description}.

Effect sizes of risk factors were randomly either protective (log odds ratio between -1.5 and -0.5) or harmful (log odds ratio between 0.5 and 1.5). The simulation also used two confounders with a prevalence of 1\%.
Essential to a contest is the scoring rule that is used to compare submissions. The scoring rule was chosen to reflect the rational of a frequentist approach, meaning that a higher emphasis is placed on keeping the false positive rate low compared to the false negative rate. Accordingly, we chose a scoring rule which penalizes false positives errors more heavily than false negative, displayed in figure \ref{fig:task_description}.
Students formed teams of up to three people. As an extra incentive, there were three statistics textbooks awarded as prizes for the winning team.

\section{Submissions}
\label{sec:submissions}

Team A used a best subset approach. They split the data into a train set (75\% of observations) and test set (25\% of observations) and fit logistic regression models with variable sizes between 3 and 7 on the train set. They chose the model which had the lowest prediction error in the test set.

Team B took a holistic approach. They first fitted penalized models with lasso and ridge regression and selected risk factors based on which models had the lowest deviances under cross-validation. Parallel to this, they graphed the number of drugs taken compared between cases and controls. Using these two pieces of information, they made an informed decision on which variables to select. The decision was influenced by a conservative attitude given that the scoring rule penalizes false positive errors heavily which lead the team to select only the 3 most obvious candidates.

Team C used the leaps algorithm \citep{Furnival1974} to perform an exhaustive search for the best subsets of the variables. They chose the model which had the lowest prediction estimated by 4-fold cross-validation.

\begin{figure}
\centering
\includegraphics[width=\textwidth]{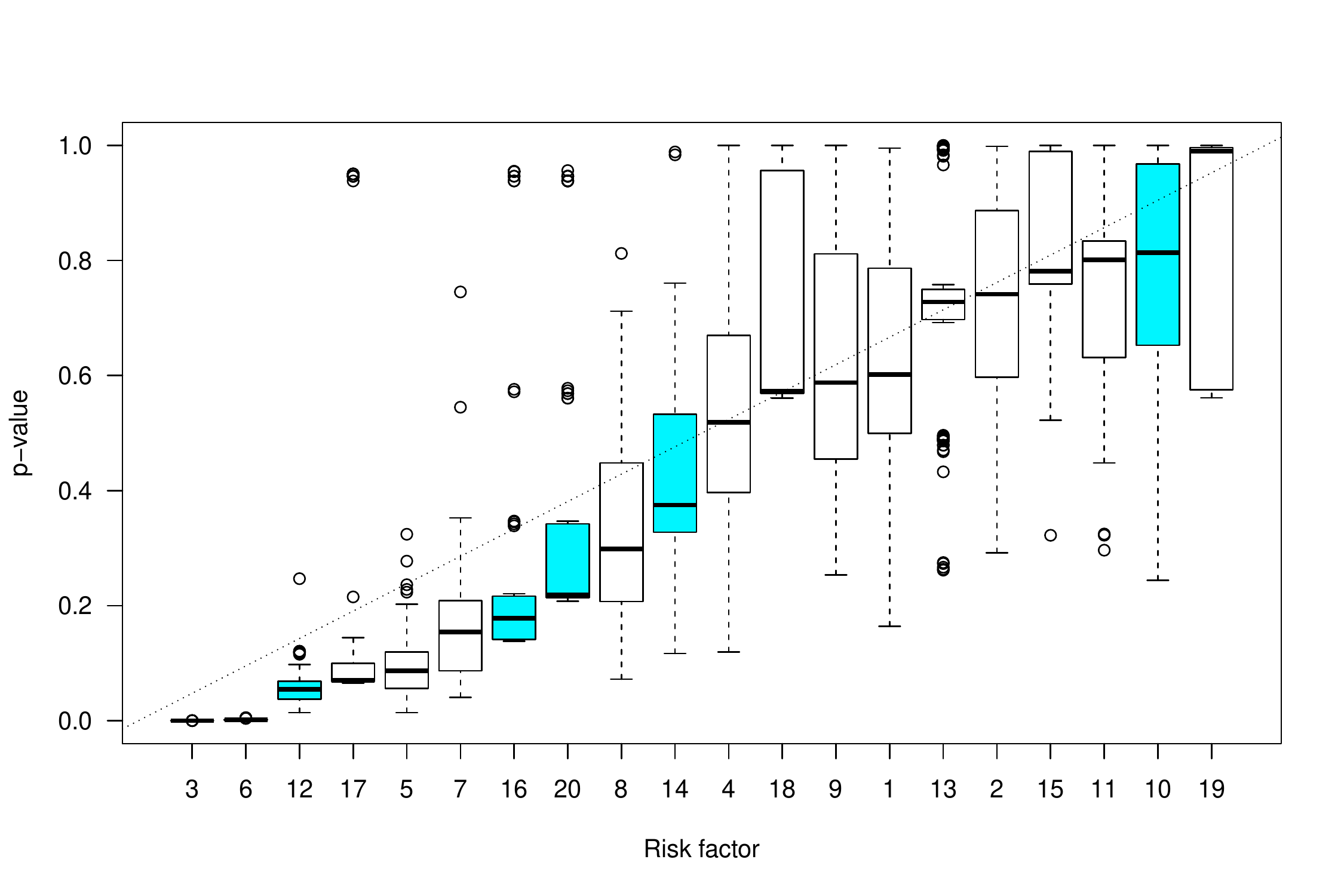}
\caption{The selection done by Team D. Shown are the boxplots for 100 p-values from a logistic regression model using resampling, sorted by their medians. The diagonal line represents the uniform distribution of p-values if no variables have an influence. From this plot, Team D decided to select risk factors 3, 6, 12, 17 and 5. The highlighted boxplots identify the relevant variables (highlighting not visible for relevant risk factors 3 and 6). \label{fig:submission_team_d}}
\end{figure}

Team D used a resampling approach, fitting logistic regression models to 100 resamples of the data and gathering the resulting p-values. Using the fact that under the null hypothesis p-values are uniformly distributed, they plotted the p-values for every risk factor and visually looked for strong deviances from the uniform distribution. See figure \ref{fig:submission_team_d} for an illustration.

Table \ref{tab:result_overview} summarizes the selections made by every team.

\begin{table}
\caption{Result overview. ``Effect'' is the regression coefficient (i.e. log odds ratio) used in simulating the outcome. As all variables were binary, their prevalences (i.e. frequencies) are also listed. The 7 relevant variables are highlighted, and the checkmarks indicate the selections by each team.
\label{tab:result_overview}}
\begin{center}
\begin{tabular}{ccccccc}
\textbf{Variable} & \textbf{Effect} & \textbf{Prevalence (\%)} & \textbf{Team A} & \textbf{Team B} & \textbf{Team C} & \textbf{Team D} \\ 
1                 &                 & 3.0                      &                 &                 &                 &                 \\
2                 &                 & 2.5                      &                 &                 &                 &                 \\
\rowcolor[HTML]{C0C0C0} 
3                 & -0.9            & 2.1                      & \checkmark           & \checkmark           & \checkmark           & \checkmark           \\
4                 &                 & 1.8                      &                 &                 &                 &                 \\
5                 &                 & 1.5                      &                 &                 & \checkmark           & \checkmark           \\
\rowcolor[HTML]{C0C0C0} 
6                 & -0.72           & 1.2                      & \checkmark           & \checkmark           & \checkmark           & \checkmark           \\
7                 &                 & 1.0                      &                 &                 &                 &                 \\
8                 &                 & 0.9                      & \checkmark           & \checkmark           &                 &                 \\
9                 &                 & 0.7                      &                 &                 &                 &                 \\
\rowcolor[HTML]{C0C0C0} 
10                & 0.53            & 0.6                      &                 &                 &                 &                 \\
11                &                 & 0.5                      &                 &                 &                 &                 \\
\rowcolor[HTML]{C0C0C0} 
12                & -1.26           & 0.4                      &                 &                 & \checkmark           & \checkmark           \\
13                &                 & 0.4                      &                 &                 &                 &                 \\
\rowcolor[HTML]{C0C0C0} 
14                & -0.64           & 0.3                      &                 &                 &                 &                 \\
15                &                 & 0.2                      &                 &                 &                 &                 \\
\rowcolor[HTML]{C0C0C0} 
16                & -0.8            & 0.2                      & \checkmark           &                 & \checkmark           &                 \\
17                &                 & 0.2                      & \checkmark           &                 & \checkmark           & \checkmark           \\
18                &                 & 0.1                      &                 &                 &                 &                 \\
19                &                 & 0.1                      &                 &                 &                 &                 \\
\rowcolor[HTML]{C0C0C0} 
20                & -1.13           & 0.1                      & \checkmark           &                 &                 &                 \\ 
\textbf{True positives}    &                 &                          & 57\%            & 29\%            & 57\%            & 43\%            \\
\textbf{True negatives}    &                 &                          & 85\%            & 92\%            & 85\%            & 85\%            \\
\textbf{Score}    & \textbf{}       & \textbf{}                & \textbf{44}     & \textbf{31}     & \textbf{44}     & \textbf{31} \\   
\end{tabular}
\end{center}
\end{table}

\section{Discussion}
\label{sec:discussion}

The contest was a rewarding and fun exercise for both the instructor and the students. The contest was designed by one of the graduate students who is working on problems of sparse data for his PhD and was thus able to provide most of the preparatory work without extra effort by the teacher. Given that the instructions were broad, teams were able to come up with very different approaches. The participants seemingly enjoyed the contest and liked being able to independently work on data without many restrictions. Although the levels of experience varied, the contest setup allowed all participants to use methods that they felt most comfortable with. Students had to explain their approaches and reasoning to the rest of the class. The way the simulated data was generated also allowed the creator to participate in the contest without knowing the correct risk factor selection.

Many scoring rules are conceivable, but all should be strictly proper scoring rules \citep{Gneiting2007}, such as the Brier score and logarithmic score for binary outcomes and continuous ranked probability scores for continuous outcomes. This ensures that a submission which corresponds to the truth scores higher than any other submission. For this contest, a simple rule was chosen in order for the scoring to be intuitive and clear to all participants. However, one weakness of the scoring rule that was noticed only afterwards was that the symmetric scoring would lead more easily to teams having the same score. An alternative would be to use Youden's index \citep{Youden1950}. For repeating our contest, we suggest a scoring scheme as proposed in the appendix table ``Proposed scoring''.

Since the dataset was created from a participant of the course, students could work in a setting similar to potential PhD projects at the university. In addition the background of the simulated data was shown, such that students felt responsible for delivering meaningful results. Since the parameters of the simulation were hidden, students were able to compete with the approaches of the creator and the instructor. Many of the students were thus eager to beat those solutions.

In courses without contests created by advanced students, we recommend to use readily available datasets. As an example, an introductory contest also carried out as part of the course, conducted similarly to the one described here, used the titanic survival dataset from Kaggle to compare classification trees, random forests and neural nets. Since the prediction variables are easy to interpret in this well-known example it encouraged students to improve their methods using their own hypotheses.

When repeating the contest game in the classroom, some points we would stress are:

\begin{enumerate}
\item Carefully choosing a scoring rule that is in line with the objective of the contest and makes it easier to identify a unique winner team.
\item Emphasize to students that their submissions have to be explainable and should be presented to the classmates.
\item Use a contest related to students' interests or research of local groups.
\end{enumerate}

Overall, the contest format was a very rewarding learning experience for both the teacher and the students, and we greatly encourage other instructors to try the same.

\section*{Acknowledgements}
\label{sec:acknowledgements}

We would like to thank the students of the course from the winter semester 2016/2017 for their efforts and participation. 

\bibliographystyle{apalike}
\bibliography{Bibliography-MM-MC}

\begin{thebibliography}{}

\bibitem[ASA, 2014]{ASA2014}
ASA (2014).
\newblock 2014 curriculum guidelines for undergraduate programs in statistical
  science.
\newblock Available at
  \url{https://www.amstat.org/asa/education/Curriculum-Guidelines-for-Undergraduate-Programs-in-Statistical-Science.aspx}.

\bibitem[Cobb, 2015]{Cobb2015}
Cobb, G. (2015).
\newblock Mere renovation is too little too late: We need to rethink our
  undergraduate curriculum from the ground up.
\newblock {\em The American Statistician}, 69(4):266--282.

\bibitem[De~Veaux et~al., 2017]{DeVeaux2017}
De~Veaux, R., Agarwal, M., Averett, M., Baumer, B., Bray, A., Bressoud, T.,
  Bryant, L., Cheng, L., Francis, A., Gould, R., Kim, A.~Y., Kretchmar, M., Lu,
  Q., Moskol, A., Nolan, D., Pelayo, R., Raleigh, S., Sethi, R.~J., Sondjaja,
  M., Tiruviluamala, N., Uhlig, P., Washington, T., Wesley, C., White, D., and
  Ye, P. (2017).
\newblock Curriculum guidelines for undergraduate programs in data science.
\newblock {\em Annual Review of Statistics and Its Application}, 4(1):15--30.

\bibitem[Furnival and Wilson, 1974]{Furnival1974}
Furnival, G.~M. and Wilson, R.~W. (1974).
\newblock Regression by leaps and bounds.
\newblock {\em Technometrics}, 16(4):499--511.

\bibitem[Gerds, 2016]{Gerds2016}
Gerds, T. (2016).
\newblock The kaplan-meier theatre.
\newblock {\em Teaching Statistics}, 38(2):45--49.

\bibitem[Gneiting and Raftery, 2007]{Gneiting2007}
Gneiting, T. and Raftery, A.~E. (2007).
\newblock Strictly proper scoring rules, prediction, and estimation.
\newblock {\em Journal of the American Statistical Association},
  102(477):359--378.

\bibitem[Goldstein, 1969]{Goldstein1969}
Goldstein, G. (1969).
\newblock A classroom approach to simulation.
\newblock {\em Mathematics Teaching}, 49:14--20.

\bibitem[Hardin et~al., 2015]{Hardin2015}
Hardin, J., Hoerl, R., Horton, N.~J., Nolan, D., Baumer, B., Hall-Holt, O.,
  Murrell, P., Peng, R., Roback, P., Temple~Lang, D., et~al. (2015).
\newblock Data science in statistics curricula: Preparing students to "think
  with data".
\newblock {\em The American Statistician}, 69(4):343--353.

\bibitem[Hastie et~al., 2009]{Hastie2009}
Hastie, T., Tibshirani, R., and Friedman, J. (2009).
\newblock {\em The Elements of Statistical Learning: Data Mining, Inference and
  Prediction}.
\newblock Springer Publishing Company, New York, NY, 2nd edition.

\bibitem[Horton and Hardin, 2015]{Horton2015}
Horton, N.~J. and Hardin, J. (2015).
\newblock Teaching the next generation of statistics students to "think with
  data": special issue on statistics and the undergraduate curriculum.
\newblock {\em The American Statistician}, 69(4):259--265.

\bibitem[James et~al., 2013]{James2013}
James, G., Witten, D., Hastie, T., and Tibshirani, R. (2013).
\newblock {\em An Introduction to Statistical Learning; with Applications in
  R}.
\newblock Springer Publishing Company, New York, NY.

\bibitem[Lloyd, 2014]{Lloyd2014}
Lloyd, J.~R. (2014).
\newblock Gefcom2012 hierarchical load forecasting: Gradient boosting machines
  and gaussian processes.
\newblock {\em International Journal of Forecasting}, 30(2):369--374.

\bibitem[Martinez and Walton, 2014]{Martinez2014}
Martinez, M.~G. and Walton, B. (2014).
\newblock The wisdom of crowds: The potential of online communities as a tool
  for data analysis.
\newblock {\em Technovation}, 34(4):203--214.

\bibitem[Narayanan et~al., 2011]{Narayanan2011}
Narayanan, A., Shi, E., and Rubinstein, B.~I. (2011).
\newblock Link prediction by de-anonymization: How we won the kaggle social
  network challenge.
\newblock In {\em Neural Networks (IJCNN), The 2011 International Joint
  Conference on}, pages 1825--1834.

\bibitem[Orriols et~al., 2010]{Orriols2010}
Orriols, L., Delorme, B., Gadegbeku, B., Tricotel, A., Contrand, B., Laumon,
  B., Salmi, L.-R., Lagarde, E., Group, C.~R., et~al. (2010).
\newblock Prescription medicines and the risk of road traffic crashes: a french
  registry-based study.
\newblock {\em PLoS Med}, 7(11):e1000366.

\bibitem[Pers et~al., 2009]{Pers2009}
Pers, T.~H., Albrechtsen, A., Holst, C., S{\o}rensen, T.~I., and Gerds, T.~A.
  (2009).
\newblock The validation and assessment of machine learning: a game of
  prediction from high-dimensional data.
\newblock {\em PLoS One}, 4(8):e6287.

\bibitem[Selkirk, 1973]{Selkirk1973}
Selkirk, K. (1973).
\newblock Random models in the classroom.
\newblock {\em Mathematics in School}, 2(6):5--6.

\bibitem[Taieb and Hyndman, 2014]{Taieb2014}
Taieb, S.~B. and Hyndman, R.~J. (2014).
\newblock A gradient boosting approach to the kaggle load forecasting
  competition.
\newblock {\em International Journal of Forecasting}, 30(2):382--394.

\bibitem[Wagaman, 2016]{Wagaman2016}
Wagaman, A. (2016).
\newblock Meeting student needs for multivariate data analysis: A case study in
  teaching an undergraduate multivariate data analysis course.
\newblock {\em The American Statistician}, 70(4):405--412.

\bibitem[Wierman, 2016]{Wierman2016}
Wierman, J.~C. (2016).
\newblock The class joke contest: Encouraging creativity and improving
  attendance.
\newblock {\em The American Statistician}, 70(3):257--259.

\bibitem[Youden, 1950]{Youden1950}
Youden, W.~J. (1950).
\newblock Index for rating diagnostic tests.
\newblock {\em Cancer}, 3(1):32--35.

\end{thebibliography}

\includepdf[pages=1]{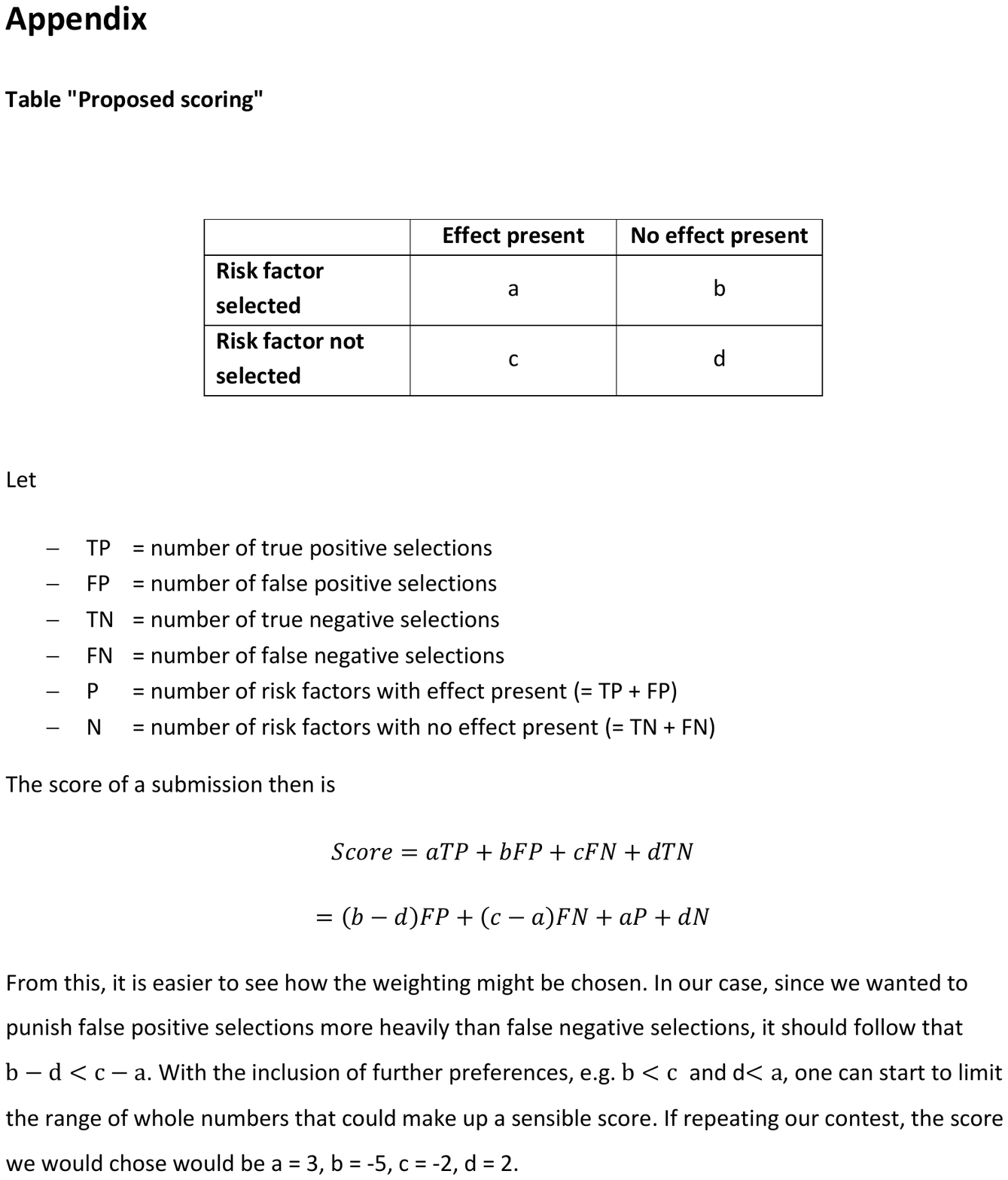}

\end{document}